\documentstyle[12pt]{article}
\newcommand{\del}{{\partial}}
\newcommand{\beq}{\begin{eqnarray}}
\newcommand{\eeq}{\end{eqnarray}}
\newcommand{\be}{\begin{eqnarray*}}
\newcommand{\ee}{\end{eqnarray*}}

\newcommand{\bx}{{\bf x}}

\newcommand{\ve}{\varepsilon}

\newcommand{\ex}[1]{\langle\,#1\,\rangle}
\newcommand{\D}{{\cal D}}
\def\bg#1{\mbox{\boldmath$#1$}}

\begin{document}
\hspace{110mm} {\it June 15, 1993}

\vspace{10mm}
\centerline{\LARGE\bf Scale anomalies in non-relativistic}
\bigskip
\centerline{\LARGE\bf field theories in 2+1 dimensions}
\vskip 10mm
\centerline{T. Haugset and  F. Ravndal}
\centerline{\it Institute of Physics}
\centerline{\it University of Oslo}
\centerline{\it N-0316 Oslo, Norway}

\vspace{10mm}
{\bf Abstract:} {\footnotesize
{}From the one-loop effective potential for a gas of
non-relativistic bosons in two spatial dimensions interacting via a
$\delta$-function
potential at zero-temperature and finite chemical potential, the anomaly of the
energy-momentum
tensor follows directly. It is also similarly derived when the bosons have an
additional
Chern-Simons interaction. In the special case of anyons, the scale anomaly
vanishes to one-loop
order in the effective potential and also to second order in the statistical
angle.

}

\vspace{10mm}
While the mass $m$ in relativistic quantum field theories is a dynamic quantity
dependent on quantum effects, it is just a passive parameter in
non-relativistic
theories and can be transformed away by using appropriate units for length.
This corresponds to taking $m = 1$. The scale dimensions of the fields and
coupling
constants in quantum units where $\hbar = 1$ then follows from the action which
is dimensionless.

Bergman has recently \cite{Bergman} considered non-relativistic	bosons
in 2+1 dimensions interacting via a $\delta$-function potential with strength
$g$. Describing them by the scalar Schr\"odinger field $\phi(t,\bx)$,
the physics of this many-particle system will then follow from the
corresponding Lagrangian density
\beq
    {\cal L} = i\phi^*\del_t\phi - {1\over 2}|{\bg\nabla}\phi|^2
             - g(\phi^*\phi)^2			        	 \label{eq.1}
\eeq
Defining now the dimension of length to be dim$(x) = -1$, it follows that
dim$(t) = - 2$. The corresponding dimensions for momentum and energy are
dim$(p) = 1$ and dim$(E) = 2$. Obviously, this is in accordance with $E =
p^2/2$
for a free particle. From the action $S = \int\!dtd^2x\,{\cal L}$
we then find that dim$({\cal L}) = 4$ and hence dim$(\phi) = 1$.
The coupling constant $g$ is therefore dimensionless in 2+1 dimensions.
Thus, the Lagrangian (\ref{eq.1}) contains no dimensional parameters.
This implies that the action is invariant
under the infinitesimal scale transformations $\delta x = \lambda x$ and
$\delta t = 2\lambda t$  which induces the corresponding change
$\delta \phi = \lambda(1 + \bx\cdot{\bg\nabla} + 2t\,\del_t)\phi$ in the field.
As pointed out by Jackiw and Pi \cite{Jackiw} it then follows that the
energy-momentum tensor $T_{\mu\nu}$ must satisfy
\beq
     2T_{00} = T_{xx} + T_{yy}                                \label{eq.2}
\eeq
Since the energy density of the particles is ${\cal E} = \ex{T_{00}}$ and their
pressure is $P = \ex{T_{xx}} = \ex{T_{yy}}$, we can also write this result
as ${\cal E} = P$. It is known to hold for free and non-relativistic bosons and
fermions in two spatial dimensions.

In relativistic quantum field theories which are invariant under scale
transformations, one can similarly show that the covariant trace $T^\mu_\mu$
of the energy-momentum tensor must vanish at the classical level. This
applies to massless particles like photons in 3+1 dimensions and gives the
well-known relation ${\cal E} = 3P$ between energy density and pressure.

Bergman \cite{Bergman} showed that point-like interaction in (\ref{eq.1})
must be regularized in order to give finite results. This introduces
a new, dimensionfull parameter in the quantum theory which breaks the scale
invariance. Using the renormalization group applied to the one-particle
irreducible vertex functions, he finds an extra, anomalous term
$g^2(\phi^*\phi)^2/2\pi$ on the right-hand side of (\ref{eq.2}).
For a system with constant particle density $\rho = \ex{\phi^*\phi}$, we
then have
\beq
     {\cal E} = P - {\rho^2\over 2\pi}g^2                     \label{eq.3}
\eeq
where now $g$ is the renormalized coupling constant.

The coefficient in front of $\rho^2$ is essentially the $\beta$-function of
the interacting theory. It arises from the one-loop contribution to the
renormalization of the coupling constant where the only divergence in the
theory
appears. Thus the above result is claimed to be exact to all orders in the
coupling constant.

Since the anomalous term is a one-loop effect, it should then also be seen
more directly in the one-loop effective potential of the theory. This has
previously been calculated by Lozano \cite{Lozano}. At its minimum, it
equals the free energy density of the system. When the system is at a given
temperature $T = 1/\beta$ and chemical potential $\mu$, it can be obtained
from the path integral
\beq
    \Xi = \int\!\D\phi^*\D\phi\,\exp{\left(-\int_0^\beta\!d\tau\!\int\!d^2x
          [\phi^*\del_{\tau}\phi + {1\over 2}|{\bg\nabla}\phi|^2
        + g(\phi^*\phi)^2 - \mu \phi^*\phi]\right)}
\eeq
for the partition function. In the exponent we now have the Euclidean action
where $\tau$ denotes imaginary time. The pressure is then $P =
\log{\Xi}/\beta V$ where $V$ is the volume of the system.

At zero temperature the classical field takes the constant values
$\phi^*\phi = \mu/2g$ where the action has a minimum. It gives the
corresponding pressure $P = \mu^2/4g$.
Including the quantum fluctuations around the classical value to lowest order,
Lozano derived the one-loop result
\beq
    P = {\mu^2\over 4g} - {\mu^2\over 16\pi}\left(1 + 2\log{\mu\over
M^2}\right)
                                                               \label{eq.4}
\eeq
The new parameter $M$ comes from the renormalization of the coupling constant.
One also has to perform a renormalization of the chemical potential.

The density of particles can now be obtained from the thermodynamic relation
$\rho = \del P/\del\mu$. It gives
\beq
    \rho = {\mu\over 2g} - {\mu\over 4\pi}\left(1 + \log{\mu\over M^2}\right)
                                                                 \label{eq.5}
\eeq
Since the energy density follows from the Legendre transformation
${\cal E} = \rho\mu - P$, the scale anomaly is simply $\rho\mu - 2P$. From
(\ref{eq.4})
and (\ref{eq.5}) we then recover immediately the Bergman result (\ref{eq.3}).

Lozano \cite{Lozano} also calculated the zero-temperature pressure of the
boson gas when it is minimally coupled to a gauge field $a_\mu(t,\bx)$ whose
dynamics is governed by the Chern-Simons term. The Lagrangian (\ref{eq.1})
is then extended to
\beq
    {\cal L} = i\phi^*(\del_t + i a_t)\phi
             - {1\over 2}|({\bg\nabla} - i\,{\bf a})\phi|^2
             + {1\over 4\theta}\ve^{\mu\nu\lambda}a_\mu\del_\nu a_\lambda
             - g(\phi^*\phi)^2			        	      \label{eq.7}
\eeq
{}From the kinetic terms we see that dim$(a_t) = 2$ while dim$(a_x)$ =
dim$(a_y) = 1$.
Thus, the Chern-Simons coupling constant $\theta$ is also dimensionless in
this non-relativistic theory. The scale invariance at the classical level is
expected to be destroyed by quantum effects and the full theory should have
a scale anomaly.

The calculation of the pressure at zero temperature is obtained as above from
the path integral for the partition function. One must add a gauge-fixing
term which is taken to define a family of Coulomb gauges. There is no
gauge dependence in the one-loop result for the pressure. It can be written as
\beq
    P = {\mu^2\over 4g} - {\mu^2\over 16\pi}\left(1 - {\theta\over g}\right)^2
      + {\mu^2\over 8\pi}\left(1 - {\theta^2\over g^2}\right)
        \log{M^2/\mu\over 1 + {\theta}/g}                     \label{eq.8}
\eeq
where $M$ is again the renormalization mass. The particle and energy densities
are obtained as above. We then find for the scale anomaly
\beq
      {\cal E} - P = {\rho^2\over 2\pi}(\theta^2 - g^2)         \label{eq.9}
\eeq
In constrast to the presumedly exact result (\ref{eq.3}), this anomaly will
in general receive additional contributions from higher loop corrections due
to the Chern-Simons interaction.

The scale anomaly is seen to vanish for the special values $g = \pm \theta$.
Lozano
\cite{Lozano} noted that the one-loop contribution to the energy density
vanished for
the special value $g = \theta$ of the contact interaction. As previously
pointed
out by Jackiw and Pi \cite{Jackiw}, the Lagrangian (\ref{eq.7}) then admits
special
self-dual soliton solutions.

The Lagrangian (\ref{eq.7}) can also be used to describe anyons \cite{Wilczek}
where $\theta$ is the statistical angle \cite{LM}. Without the contact term one
finds
divergences in the wavefunctions even in the two-anyon system when the
particles get
very close \cite{GHKL}. Similarly, one needs a repulsive $\delta$-function
potential with
the specific strength $g = \theta$ between each pair of particles in order to
get finite
results for many-anyon systems from first-order perturbation theory about the
boson point
$\theta = 0$ \cite{Sen.1}. In addition, Valle Basagoiti \cite{Valle} has shown
that for
the same value of the contact interaction the divergencies
in the calculation of the partition function of the anyon gas from the
Lagrangian
(\ref{eq.7}) exactly cancels to order $\theta^2$ and to second order in the
fugacity
$z = \exp({\beta\mu})$. The finite result agrees with the known second virial
coefficient for the anyon gas \cite{ASWZ}.

With this special value of the contact interaction, we see from (\ref{eq.9})
that the
scale anomaly for the anyon gas vanishes at the one-loop level we consider
here.
The pressure (\ref{eq.8}) is then simply $P= \mu^2/4\theta$. In terms of the
density
$\rho = \mu/2\theta$ it is $P = \theta\rho^2$. The pressure in a free gas of
bosons with
$\theta = 0$ now comes out correctly with the value $P = 0$. Higher loop
contributions to
the effective potential will then give a more general result $P =
f(\theta)\rho^2$ where
the function $f(\theta)$ has period $2\pi$.

Recently, Emparan and Valle Basagoiti \cite{EVB} have calculated the
finite-tempe\-rature pressure in the anyon gas exactly to second order in
$\theta$ and to all orders in the fugacity with the anyonic value for the
contact coupling constant. They find that the pressure varies with the
temperature $T$ as $P = T^2F(z)$ where $F(z)$ is a function of the
fugacity. Extracting the six lowest virial coefficients, they agree with the
results of Ouvry {\it et al} \cite{Ouvry} obtained from first-quantized anyon
theory.
The corresponding energy density is ${\cal E} = - (\del\beta P/\del\beta)_z =
P$ and
thus gives a vanishing scale anomaly. Emparan and Valle Basagoiti
have also performed the same calculation around the fermion point $\theta =
\pi$ to
second order in $\theta - \pi$. There is then no need for the contact
interaction which is
now effectively replaced by the Pauli repulsion. Again there is no anomaly.
These
calculations involve three-loop Feynman diagrams and it is tempting to conclude
that the
lack of a scale anomaly holds to all orders in the coupling constant $\theta$.

The reason for this surprising and simple result is the exact cancelling of
divergences in
the anyonic theory when $g = \theta$. There is then no need for renormalization
of
the coupling constant and there will be no renormalization mass $M$ in the
thermodynamic
pressure.  In our units the scale dimension of the pressure is dim$(P) = 4$ and
temperature has the same dimension as energy, i.e. dim$(T) = 2$. We must then
have $P = T^2F(z)$ for dimensional reasons since there are no other quantities
than
the temperatur $T$ which sets the scale. There is thus no scale anomaly for the
anyon
gas.

Since the statistical angle $\theta$ does not suffer any renormalization, the
corresponding
$\beta$-function must also vanish. It shows that the quantum properties of the
anyon gas are
in some way simpler than expected \cite{Ulf}. Quite possibly the system has
symmetries which
are not yet fully understood.
Jackiw and Pi \cite{Jackiw} have noted that for this special value of the
coupling
constant $g$ the contact term is equivalent to a Pauli spin interaction which
might
indicate some non-linear realization of supersymmetry \cite{LLM} in the anyon
system.
Sen \cite{Sen.2} has also demonstrated supersymmetric features in the spectra
of
many-anyon systems.

\vspace{10mm}
We thank T.H. Hansson, J.M. Leinaas, K. Olaussen and S. Viefers for informative
discussions.

\end{document}